\documentclass[floats,aps]{revtex4}
\usepackage{calc}
\usepackage{graphicx}
\begin{document}
\def \beq{\begin{equation}}
\def \eeq{\end{equation}}
\begin{abstract}
We argue that  the naively expected singularities of the Fermi surface,
 in the mixed composite boson-composite fermion states proposed [S.H. Simon {\it et al.}, 
PRL {\bf 91}, 046803 (2003)] for the evolution of $\nu =1$ bilayer quantum Hall system with distance, are obliterated. Our conclusion is based on a careful analysis of the momentum distribution in $\nu = \frac{1}{2}$ single-layer composite-
fermion state. We point out to a possibility of the phenomenon hitherto unknown outside
Kondo lattice systems when, in a translationally invariant system,
Fermi-liquid-like portion of electrons enlarges its volume.
\end{abstract}
\title{Status of the Fermi surface in mixed composite boson - composite fermion quantum Hall states}
\author{Milica V. Milovanovi\'{c}}
\address{Institute of Physics, P.O.Box 68, 11080 Belgrade, Serbia and Montenegro}
\date{\today}
\maketitle
\vskip2pc]
\narrowtext
The nature and physics of the transition in the bilayer $\nu = 1$ 
quantum Hall (QH) system \cite{halp} between the well-established phases:
one characteristic for the distances between the layers of the order
of or smaller than magnetic length, sometimes described as ``111'' state,
and the other for larger distances, described by two separate
Fermi-liquid-like states of composite fermions (CFs) attracted recently
the attention of experimentalists \cite{kelog} and is the focus of several
theoretical papers \cite{sch,stha,srm,kimv}. Only
references \cite{stha} and \cite{srm} make a prediction
 for a coexistence region between two phases, with a unique
property, semicircle law for the longitudinal and Hall drag
resistance that was revealed in the experiments \cite{kelog}.
The reference \cite{srm}
introduces a form of the ground state of the system that may
continously interpolate between the 111 state, usually described
by composite bosons (CBs), and the two seperate Fermi-liquid-like states of
CFs. The ground state proved to be a good variational ansatz when compared
with the exact solution in numerical studies \cite{srm}. The form of
the variational state for certain distance between the layers may be
  described as one in which classically speaking
some of the electrons are in the 111 state (they make CBs) and the
others participate in two Fermi seas of CFs. Gradually the
number of CFs increases as the distance becomes larger. Therefore
the description easily accounts for the continous nature of the transition
as observed in the experiments \cite{kelog}. On the other hand the
proposal that came first, based on a phase separated picture, \cite{stha},
in which percolating puddles of one phase are in the other, well
enough exibits the transport properties measured in the experiments.
The advantage of the homogenous model (\cite{srm}), which accounts
for the same transport properties, is that it  also accounts
for the strong 111 (interlayer) correlations that occur even
deep in the CF region \cite{kelog}.

Here we study the Fermi surface singularities in the proposed wave
functions \cite{srm}. Naively they are expected at the Fermi momenta 
directly related to the number of CFs in the particular partition
of the overall number of electrons into CFs and CBs. The analysis
begins with a careful study of the $\nu = \frac{1}{2}$ CF problem,
so that the relationships found can be readily applied to the 
mixed state case. We found that the CF momentum distribution near
the naively expected Fermi momenta depend analytically on the
distance to the Fermi momenta, therefore showing no signature of the
Fermi surfaces.

Soon after Halperin, Lee, and Read \cite{hlr} proposed their theory for $\nu = \frac{1}{2}$ fractional QH effect  Bares and Wen \cite{bw} considered fermions in low dimensions interacting via a long range $ - \frac{2 \pi}{|\vec{q}|^{2}}$ interaction. They used as a good ansatz for the ground state, a wave function of the Feenberg-Jastrow type,
\beq
\Psi_{o}(\{x\}) = \prod_{i<j} |x_{i} - x_{j}|^{m} \Psi_{FS}(\{x\}),
\label{fjs}
\eeq
where $\Psi_{FS}$ denotes a Slater determinant of filled Fermi sea of free single-particle states. If $m = 2$ this construction is the Rezayi-Read \cite{rr} ground state, in the representation of CFs and when the projection to the lowest Landau level (LLL) is neglected, found to correctly captures the physics at $\nu = \frac{1}{2}$. By doing a calculation of a random phase approximation (RPA) type on (\ref{fjs}) Bares and Wen found that the leading singularity of the momentum distribution near $ k_{F}$, in two dimensions, is
\beq
\label{ls}
\delta n_{k} \approx \frac{m}{2} \{n_{k}^{o} \ln|\delta k| -
(1 - n_{k}^{o}) \ln|\delta k|\},
\eeq
where $\delta k = |\vec{k}| - k_{F}$ and $n_{k}^{o}$ denotes the free-Fermi-gas momentum distribution. They also remarked that if we interpret the rhs of (\ref{ls}) as the first term in an expansion in powers of $m$ we can write (near $k_{F}$)
\beq
n_{k} = \frac{1}{2} + \frac{1}{2} \{n_{k}^{o} |k - k_{F}|^{\frac{m}{2}} - (1 - n_{k}^{o}) |k - k_{F}|^{\frac{m}{2}} \}.
\label{extrap}
\eeq
What they did not emphasize is that if $m = 2$ and although we have a 
Luttinger-liquid type of expansion near $k_{F}$ \cite{ml} there is no nonanalytic behavior due to the odd power of $ |k - k_{F}|$ and all trace of the Fermi surface has been eliminated.

\begin{figure}
\includegraphics{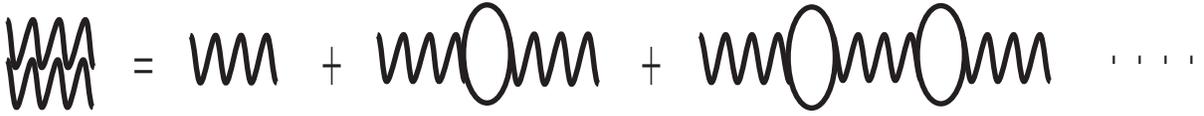}
\caption{Effective screened interaction}
\label{one}
\end{figure}

We can come to the same expressions employing the weakly-screening plasma analogy \cite{meth}, which in considering  quantum-mechanical expectations in the state, Eq.(\ref{fjs}), mimics Laughlin's plasma approach \cite{la}. In the Laughlin approach there is the perfect screening of the classical Coulomb plasma, when interaction
$- \frac{2 \pi}{|\vec{k}|^{2}}$ becomes screened as
\beq
\label{screenint}
\frac{-\frac{2 \pi}{|\vec{k}|^{2}}}{1 + \frac{2 \pi \beta m^{2}}{|\vec{k}|^{2}} s_{o}(k)},
\eeq
where $m$ is from $\nu = \frac{1}{m}$, filling factor; $\beta = \frac{2}{m}$ is the plasma inverse temperature, and $s_{o}(k)$ is the static structure factor of the noninteracting particles, in this case bosons, so that $s_{o}(k) = \rho$ - particle density, and hence  a perfect screening. More precisely it can be found \cite{thesis} that the expansion in small $m$ of  classical statistic averages (to which quantum expectations correspond) is well defined, gives the results that can be found by other methods, and allows continuation to larger than $m = 1$ values. In this context the screening is captured by the accustomed infinite summation of a geometric series described by Eq.(\ref{screenint}) and symbolically can be represented by the sum of diagrams as in Fig. \ref{one}

In the case of the weakly-screening plasma analogy due to the presence of the free-fermion Slater determinant in (\ref{fjs}), the first summation, (\ref{screenint}), that is done while organizing diagrams, gets modified, having for $s_{o}(k)$ the static structure factor of free fermion gas, which in two dimensions for small $k$ can be found to be $s_{o}^{f}(k) = \frac{3}{4} \frac{k_{F}}{\pi^{2}} k $.
This leads to not so perfect screening of the long-range interaction which becomes as $\frac{1}{r}$ instead of $\ln r $ in real space. The approach introduced parallels the RPA calculation in Ref. \cite{bw} in getting (\ref{ls}) when Fig. 1 corresponds to an RPA summation with the value of the bubble equal to $ s_{o}^{f}(k) \beta m^{2}$.

\begin{figure}
\includegraphics{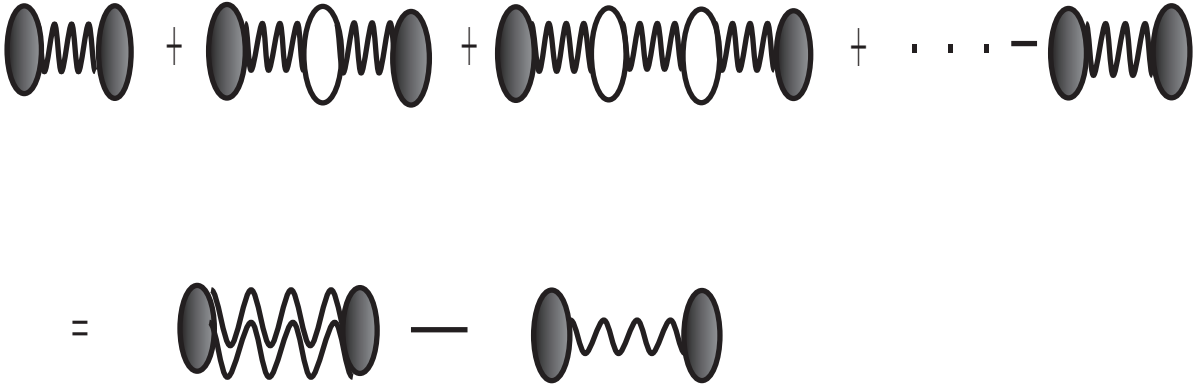}
\caption{Effective diagram contribution}
\label{two}
\end{figure}

We want to see in more detail how the equal-time CF propagator can be found, and, possibly, which additional diagrams in its calculation would lead to the conjectured expression for the CF occupation number. It is instructive to first consider how we can get the equal-time CB correlator i.e. Girvin - MacDonald correlations \cite{gmd} in the Laughlin case using the diagramatic expansion \cite{thesis}. As introduced by Girvin and MacDonald we in fact in the plasma language have to deal with two impurities of charge $\frac{m}{2}$ each, which do not interact directly. Therefore we have
\beq
\label{bplasma}
G_{B}(z,z^{'}) \sim |z - z^{'}|^{-\frac{m}{2}} \frac{Z(z,z^{'})}{Z(z,z)},
\eeq
where $Z(z,z^{'})$ is the partition function of the classical 2D plasma with inverse temperature $\beta = \frac{2}{m}$, each particle with charge $m$, as before, and two impurities with charge $\frac{m}{2}$ each at the locations $z$ and $z^{'}$. ($Z(z,z)$ is the partition function with one impurity of charge $m$ at an arbitrary location because the value of the partition function does not depend on $z$.) What we expect is that the ratio will have the following form,
\beq
\label{form}
\frac{Z(z,z^{'})}{Z(z,z)} = \exp\{- \beta \Delta f(z,z^{'})\},
\eeq
where $\Delta f(z,z^{'})$ represents the difference in the free energy between the two configurations. Indeed we can find doing the simple expansion in $m$ that the term right after the first term (of value one) is
\beq
\label{veff}
V_{eff}(|z - z^{'}|) = (\frac{m}{2})^{2} \; \int \frac{d^{2}k}{(2 \pi)^{2}} \exp\{i \vec{k}(\vec{r} - \vec{r}^{'})\}
\frac{-\frac{2 \pi}{|\vec{k}|^{2}}}{1 + \frac{2 \pi \beta m^{2}}{|\vec{k}|^{2}} \rho},
\eeq
$(\rho = \frac{1}{2 \pi m})$, which represents an effective screened interaction between two impurities and extract to mimic (\ref{form}), contributions of disconnected $V_{eff}(|z - z^{'}|)$ parts that follow so that for the final expression we get
\beq
\label{expr}
\frac{Z(z,z^{'})}{Z(z,z)} = \exp\{V_{eff}(|z - z^{'}|) \}.
\eeq
Therefore we can conclude that in calculating $G_{B}(z,z^{'})$ we have to exponentiate the value of the diagram shown in Fig. \ref{two},
and get, due to the screening, the famous algebraic decay.

Similarly, applying the same type of approximation we can get in the CF case
\beq
G_{F}(z,z^{'}) \sim G_{F}^{o}(z,z^{'}) \; \exp \{(\frac{m}{2})^{2}
\; \int \frac{d^{2}k}{(2 \pi)^{2}}
( \frac{2 \pi \beta}{|\vec{k}|^{2}}-
\frac{2 \pi \beta}{|\vec{k}|^{2}+ 2 \pi \beta m^{2} s_{o}^{f}(k)})
\exp\{i \vec{k} (\vec{r}-\vec{r}^{'})\},
\label{thgreen}
\eeq
where (the screening bubble is proportional to the static structure factor of free fermions and) $G_{F}^{o}(z,z^{'}) = \sum_{|\vec{k}|< k_{F}} e^{i \vec{k}(\vec{r}-\vec{r}^{'})}$ is the equal-time correlator of free Fermi gas. To fix the normalization we demand that the total number of CFs is the same as of noninteracting particles so that
\beq
\sum_{\vec{k}}\; \int d^{2}r \; G_{F}^{o}(\vec{r}) \; e^{- i \vec{k} \vec{r}} = N =
\sum_{\vec{k}}\; \int d^{2}r \; G_{F}(\vec{r}) \; e^{- i \vec{k} \vec{r}},
\label{equal}
\eeq
and $G_{F}^{o}(0) = G_{F}(0)$ follows. Therefore
\beq
G_{F}(z,z^{'}) = G_{F}^{o}(z,z^{'}) \; \exp \{(\frac{m}{2})^{2}
\; \int \frac{d^{2}k}{(2 \pi)^{2}}
( \frac{2 \pi \beta}{|\vec{k}|^{2}}-
\frac{2 \pi \beta}{|\vec{k}|^{2}+ 2 \pi \beta m^{2} s_{o}^{f}(k)})
[\exp\{i \vec{k} (\vec{r}-\vec{r}^{'})\}-1]\}.
\label{finalex}
\eeq
 
And indeed by taking $n_{k} = \int e^{-i \vec{k} \vec{r}} G_{F}(\vec{r})$ and considering the first nontrivial contribution in the expansion of the exponential in (\ref{finalex}) we get
\beq
\delta n_{k} = (\frac{m}{2})^{2}\; \int  \frac{d^{2}q}{(2 \pi)^{2}}
( \frac{2 \pi \beta}{|\vec{q}|^{2}}-
\frac{2 \pi \beta}{|\vec{q}|^{2}+ 2 \pi \beta m^{2} s_{o}^{f}(q)})
[n_{\vec{k}-\vec{q}}^{o} (1 - n_{k}^{o})-
n_{k}^{o}(1 - n_{\vec{k}-\vec{q}}^{o})],
\label{occbw}
\eeq
exactly the same expression as Eq.(86) of Ref. \cite{bw}. Once we specify that $k$ is near $k_{F}$, assume a flat Fermi surface and neglect the contribution of the (weakly) screened interaction in (\ref{occbw}) we can get, as in \cite{bw}, the leading singularity in $n_{k}$ given by Eq.(\ref{ls}).

But unfortunately after a suitable regrouping of free-Fermi-gas occupation numbers we can prove that the second nontrivial contribution in the expansion of the exponential (with the neglect of the screened interaction) is equal to zero. That does not mean that the conjectured contribution (Eq.(\ref{extrap})) is absent. Here we have very likely the situation that due to the nonanlytic nature of the attempted expansion in the CF case we can not generate corrections to the terms linear in $m$. That conclusion supports also the finding \cite{thesis} that when the same expansion was applied in the calculation of the static structure factor of the CF state a first correction to the RPA result could not be generated although it was expected on the grounds that the correction would have made the infered LLL-projected static structure factor positive definite what by its definition it should be.

Therefore, very likely the exponential prescription (used to get Eq.(\ref{extrap})) is a
valid one although there is no straightforward expansion to prove it. Once we accept the prescription
we are left to wonder where is the expected nonanalyticity at $ \nu = \frac{1}{m}, \frac{m}{2} =
 odd \; integer$ (see Eq.(\ref{extrap}) for that case).
A trace of the real Fermi-surface nonanalyticity  at $\frac{m}{2}= odd \; integer$ may be seen in the expansion only if we take into account the screened interaction (second part) in Eq.(\ref{occbw}). As an effective contribution from this part we have
\beq
\delta n_{k} = \frac{1}{2 \pi m s_{o}} [ |\delta k|
\ln|\delta k|\; n^{o}(k) - (1 - n^{o}(k))\;
|\delta k|
\ln|\delta k|],
\label{nona}
\eeq
where $s_{o} = \frac{3}{4} \frac{k_{F}}{\pi^{2}}$.
%, although certainly due to the impossibility of finding out further contributions this can not be a final expression with fixed, final coefficients for the nonanalytic behavior of the occupation number near the Fermi surface.
If we apply the exponential prescription again, taking also into account this second
contribution, we have, with $ \frac{1}{2 \pi m s_{o}} \equiv c$ and, for $\frac{m}{2} = 1$,
for the contribution in the vicinity of $k_{F}$,
\begin{equation}
n_{k} = \frac{1}{2} + \frac{1}{2} [ |\delta k| \exp\{c |\delta k| \ln|\delta k|\} \; n^{o}(k)
- ( 1 - n^{o}(k)) |\delta k| \exp\{c |\delta k| \ln|\delta k|\} ].
\label{fin}
\end{equation}
Here a (weak) nonanalyticity is retained. Namely, in Eq.(\ref{fin}) we have singular (at $k_{F}$) the
second derivative of $ |\delta k| \exp\{c |\delta k| \ln|\delta k|\} $ with respect to $|\delta k|$.
Therefore, a trace of the Fermi surface at $ \nu = \frac{1}{m} , \frac{m}{2}=1 $, is present
because of the found nonanalytic behavior. (Such a behavior exists also for $ \frac{m}{2} =
 odd > 1$ but is weaker having singular higher derivatives.)

In the following we will give an example where aforementioned mechanism for getting the Fermi surface (nonanalyticity) does not work due to  strong correlations of the CFs with other particles of the system. This is the case of the mixed CB - CF quantum Hall states proposed in \cite{srm}
to describe the evolution of the bilayer $\nu = 1$ QH system with distance between layers.

If we neglect the LLL projection again and assume that for our purposes we can also neglect the overall antisymmetrization between CB and CF parts that makes the mixed state completely antisymmetric and an electronic wave function, we can write it in the quasiparticle representation as
\begin{eqnarray}
&\Psi_{o}(z_{\uparrow},z_{\downarrow},w_{\uparrow},w_{\downarrow})= \;\; \; \; \; \; \; \; \; & \nonumber \\
& \prod_{i<j} |z_{i \uparrow} - z_{j \uparrow}|^{n}
\prod_{k<l} |z_{k \downarrow} - z_{l \downarrow}|^{n}
\prod_{p<q} |z_{p \uparrow} - z_{q \downarrow}|^{n} & \nonumber \\
& \prod_{i,j} |z_{i \uparrow} - w_{j \uparrow}|^{n}
\prod_{k,l} |z_{k \uparrow} - w_{l \downarrow}|^{n}
\prod_{p,q} |z_{p \downarrow} - w_{q \uparrow}|^{n}
\prod_{r,s} |z_{r \downarrow} - w_{s \downarrow}|^{n} & \nonumber \\
& \prod_{i<j} |w_{i \uparrow} - w_{j \uparrow}|^{m}
\Psi_{FS}(\{w_{\uparrow}\})
\prod_{k<l} |w_{k \downarrow} - w_{l \downarrow}|^{m}
\Psi_{FS}(\{w_{\downarrow}\}), &
\label{bofe}
\end{eqnarray}
where  $z_{\uparrow}$ and $z_{\downarrow}$ denote CB coordinates and $ w_{\uparrow}$ and $w_{\downarrow}$ denote CF coordinates with arrows specifying to which layer quasiparticles belong. The total numbers of bosons are equal as well the total numbers of fermions, and $n = \frac{m}{2} = $ odd integer.

We want to find out (the asymptotic behavior of ) the equal-time
correlator of a CF (belonging to one of the layers). It is not hard to conclude that in this case with assumptions similar to the ones done in the single-layer case, we get $G_{F}(w,w^{'})$ by simply taking for the value of the ``polarization'' bubble -
\beq
\beta m^{2} s_{o}^{f}(k) + \beta n^{2} \rho_{b},
\label{sub}
\eeq
instead of $\beta m^{2} s_{o}^{f}(k)$ only in Eq.(\ref{finalex}), where $\rho_{b}$ denotes the total (up plus down) density of bosons and in $s_{o}^{f}(k)$ we have to take $k_{F} = \sqrt{4 \pi \rho_{f}}$ where $\rho_{f}$ is the density of fermions of one layer only. In this case we work with a completely screened interaction between the two impurities which does not produce nonanalytic contributions.

Therefore we find that at the total fillings of bilayer at which we can expect bose-fermi mixed states, $\nu = \frac{2}{m}; \; m = 2,6,\ldots$ the naively expected Fermi surface(s)
 can not exist due to our analysis. 
This outcome remind us of the similar disappearance of the small
(naively) expected Fermi momentum in the Kondo lattice systems
\cite{yoa,osh} due to the Luttinger theorem \cite{lutt}.
In the case considered in the paper we do not know for sure if we 
deal with (overall) Fermi-liquid-like states and a complete analogy
(in which CBs and CFs play the roles of
of localized spin $1/2$ local moments and conduction electrons respectively) 
is still missing. Further insights into the physics of the mixed states are
necessary.
%In a broader sense this conclusion 
%also complies with the Luttinger theorem for interacting fermions \cite{lutt}
%similarly to the generalizations in \cite{yoa} and \cite{osh} in the
% Kondo lattice case. In this case of CFs and CBs (instead of
%conduction electrons and localized spin $1/2$ local moments)
%we generalize, and view the mixed states as Fermi-liquid-like states with
%necessarilly large Fermi volume, evolving from two decoupled 
%Fermi-liquid-like states with the gradual inclusion of
%interlayer interaction that preserves the Fermi volume, so that
%at the transition from pure (111) state to the mixed region there is a jump
%in the Fermi volume. The transition is still continuous as we characterize
%the ordering in the CF region as CB $p$-wave pairing \cite{mvm}, 
%so that the number of CFs is still meaningful number continously
%changing from zero at the transition point to the number of electrons
%when the layers are far apart \cite{pivo}.

The author thanks Ashvin Vishwanath for asking the question discussed in the paper,
and Aspen Center for Physics for its hospitality. This work was supported by
Grant No. 1899 of the Serbian Ministry of Science, Technology, and Development.

\end{document}